\documentclass[12pt]{article}
\usepackage{feynmf,amssymb,psfig,fullpage}
\begin{document}

%%%%%%%%%%%%%%%%%%%%%%%%%%%%%%%%%%%%%%%%%%%%%%%%%%%%%%%%%%
%%%%%%%%%%%%%% Feynman Diagrams Definitions %%%%%%%%%%%%%
%%%%%%%%%%%%%%%%%%%%%%%%%%%%%%%%%%%%%%%%%%%%%%%%%%%%%%%%%%
% Time-stamp: <1999-10-13 19:26:31 vieyra>
%
\unitlength=1mm
\newcommand{\choneladder}{\hspace{1mm}
% \parbox[pos]{width}{text}
     \parbox{14mm}{ 
      \begin{fmfgraph}(10,20)
             \fmfset{arrow_len}{2.5mm}
             \fmfset{wiggly_len}{1.7mm}
             \fmfset{dot_size}{0.8thick}
             \fmfleft{i1,i2} \fmfright{o1,o2}
             \fmfstraight
             \fmfleft{v1,v2,v3,v4,v5}
             \fmfright{v6,v7,v8,v9,v10} 
                        \fmf{fermion}{v1,v2,v7,v8,v3,v4,v9,v10}
                        \fmf{photon}{v2,v3}\fmf{photon}{v8,v9}
                        \fmffreeze
                        \fmf{photon}{v6,v7}
                        \fmf{photon}{v4,v5}                       
             \fmfdot{v2,v3,v4,v7,v8,v9}
      \end{fmfgraph}
      }   
 }
%%%%%%%%%%%%%%%%%%%%%%%%%%%%%%%%%%%%%%%%%%%%%%%%%%%%%%%%%%%%%
\newcommand{\choneladderc}{\hspace{1mm}
   \parbox{14mm}{
      \begin{fmfgraph}(10,20)
             \fmfset{arrow_len}{2.5mm}
             \fmfset{wiggly_len}{1.7mm}
             \fmfset{dot_size}{0.8thick}
             \fmfleft{i1,i2} \fmfright{o1,o2}
                        \fmf{fermion,tension=1.7}{v3,o2}
                        \fmf{photon}{v3,i2}
                        \fmf{photon,tension=1.7}{o1,v1}
                        \fmf{fermion}{i1,v1}
                        \fmf{phantom}{i1,v1,v2,v3,i2}
                        \fmf{photon,left}{v1,v2}
                        \fmf{fermion,right}{v1,v2}
                        \fmf{fermion,left}{v2,v3}
                        \fmf{photon,right}{v2,v3}
             \fmfdot{v1,v2,v3}
      \end{fmfgraph} 
      }
    }
%%%%%%%%%%%%%%%%%%%%%%%%%%%%%%%%%%%%%%%%%%%%%%%%%%%%%%%%%%%%%
\newcommand{\chzeroneg}{\hspace{1mm}
   \parbox{14mm}{ 
      \begin{fmfgraph}(10,20)
             \fmfset{arrow_len}{2.5mm}
             \fmfset{wiggly_len}{1.7mm}
             \fmfset{dot_size}{0.8thick}
             \fmfleft{i1,i2} \fmfright{o1,o2}
             \fmfstraight
             \fmfleft{v1,v2,v3,v4,v5}
             \fmfright{v6,v7,v8,v9,v10} 
                  \fmf{fermion}{v1,v2,v3,v4,v5}
                  \fmf{fermion}{v10,v9,v8,v7,v6}
                  \fmf{photon}{v2,v7}\fmf{photon}{v3,v8}\fmf{photon}{v4,v9}
             \fmfdot{v2,v3,v4,v7,v8,v9}
      \end{fmfgraph}
      }
    }
%%%%%%%%%%%%%%%%%%%%%%%%%%%%%%%%%%%%%%%%%%%%%%%%%%%%%%%%%%%%%
\newcommand{\chzeroposA}{\hspace{1mm}
   \parbox{14mm}{ 
      \begin{fmfgraph}(10,20)
             \fmfset{arrow_len}{2.5mm}
             \fmfset{wiggly_len}{1.7mm}
             \fmfset{dot_size}{0.8thick}
             \fmfleft{i1,i2} \fmfright{o1,o2}
             \fmfstraight
             \fmfleft{v1,v2,v3,v4,v5}
             \fmfright{v6,v7,v8,v9,v10} 
                        \fmf{fermion}{v1,v2,v3,v8,v7,v6}
                        \fmf{fermion}{v10,v9,v4,v5}
                        \fmf{photon}{v2,v7}\fmf{photon}{v3,v4}
                        \fmf{photon}{v8,v9}
             \fmfdot{v2,v3,v4,v7,v8,v9}
      \end{fmfgraph}
      }
    }               
%%%%%%%%%%%%%%%%%%%%%%%%%%%%%%%%%%%%%%%%%%%%%%%%%%%%%%%%%%%%%
\newcommand{\chzeroposB}{\hspace{1mm}
   \parbox{14mm}{
      \begin{fmfgraph}(10,20)
             \fmfset{arrow_len}{2.5mm}
             \fmfset{wiggly_len}{1.7mm}
             \fmfset{dot_size}{0.8thick}
             \fmfleft{i1,i2} \fmfright{o1,o2}
             \fmfstraight
             \fmfleft{v1,v2,v3,v4,v5}
             \fmfright{v6,v7,v8,v9,v10} 
                        \fmf{fermion}{v3,v4,v9,v8,v3}
                        \fmf{fermion}{v1,v2,v7,v6}
                        \fmf{photon}{v2,v3}\fmf{photon}{v7,v8}
                        \fmf{photon}{v4,v5}\fmf{photon}{v9,v10}
             \fmfdot{v2,v3,v4,v7,v8,v9}
      \end{fmfgraph}
      }
  }
%%%%%%%%%%%%%%%%%%%%%%%%%%%%%%%%%%%%%%%%%%%%%%%%%%%%%%%%%%%%%
\newcommand{\chzeroposC}{\hspace{1mm}
   \parbox{14mm}{
      \begin{fmfgraph}(10,30)
             \fmfset{arrow_len}{2.5mm}
             \fmfset{wiggly_len}{1.7mm}
             \fmfset{dot_size}{0.8thick}
             \fmfleft{i1,i2} \fmfright{o1,o2}
             \fmfstraight
             \fmfleft{v1,v2,v3,v4,v5,v6}
             \fmfright{v7,v8,v9,v10,v11,v12} 
                        \fmf{fermion}{v2,v3,v9,v8,v2}
                        \fmf{fermion}{v5,v4,v10,v11,v5}
                        \fmf{photon}{v1,v2}\fmf{photon}{v7,v8}
                        \fmf{photon}{v3,v4}\fmf{photon}{v9,v10}
                        \fmf{photon}{v5,v6}\fmf{photon}{v11,v12}
             \fmfdot{v2,v3,v4,v5,v8,v9,v10,v11}
      \end{fmfgraph}
      }
  }
%%%%%%%%%%%%%%%%%%%%%%%%%%%%%%%%%%%%%%%%%%%%%%%%%%%%%%%%%%%%%
\newcommand{\chzeroposD}{\hspace{1mm}
   \parbox{14mm}{
      \begin{fmfgraph}(10,30)
             \fmfset{arrow_len}{2.5mm}
             \fmfset{wiggly_len}{1.7mm}
             \fmfset{dot_size}{0.8thick}
             \fmfleft{i1,i2} \fmfright{o1,o2}
             \fmfstraight
             \fmfleft{v1,v2,v3,v4,v5,v6}
             \fmfright{v7,v8,v9,v10,v11,v12} 
                        \fmf{fermion}{v3,v4,v10,v9,v3}
                        \fmf{fermion}{v1,v2,v8,v7}
                        \fmf{fermion}{v12,v11,v5,v6}
                        \fmf{photon}{v2,v3}\fmf{photon}{v8,v9}
                        \fmf{photon}{v4,v5}\fmf{photon}{v10,v11}
             \fmfdot{v2,v3,v4,v5,v8,v9,v10,v11}
      \end{fmfgraph}
      }
    }
%%%%%%%%%%%%%%%%%%%%%%%%%%%%%%%%%%%%%%%%%%%%%%%%%%%%%%%%%%%%%
\newcommand{\chzeroposE}{\hspace{1mm}
   \parbox{14mm}{ 
      \begin{fmfgraph}(10,30)
             \fmfset{arrow_len}{2.5mm}
             \fmfset{wiggly_len}{1.7mm}
             \fmfset{dot_size}{0.8thick}
             \fmfleft{i1,i2} \fmfright{o1,o2}
             \fmfstraight
             \fmfleft{v1,v2,v3,v4,v5,v6}
             \fmfright{v7,v8,v9,v10,v11,v12} 
                        \fmf{fermion}{v1,v2,v8,v7}
                        \fmf{fermion}{v12,v11,v10,v9,v3,v4,v5,v6}
                        \fmf{photon}{v4,v10}\fmf{photon}{v5,v11}
                        \fmf{photon}{v2,v3}\fmf{photon}{v8,v9}
             \fmfdot{v2,v3,v4,v5,v8,v9,v10,v11}
      \end{fmfgraph}
      }
    }
%%%%%%%%%%%%%%%%%%%%%%%%%%%%%%%%%%%%%%%%%%%%%%%%%%%%%%%%%%%%%
\newcommand{\chonetree}{\hspace{1mm}
   \parbox{18mm}{ 
      \begin{fmfgraph*}(14,10)
             \fmfset{arrow_len}{2.5mm}
             \fmfset{wiggly_len}{1.7mm}
              \fmfset{dot_size}{0.8thick}\fmfdotn{v}{2}
             \fmfleft{i1,i2} \fmfright{o1,o2}
  \fmf{fermion}{v1,v2}
  \fmf{fermion}{i1,v1}
  \fmf{fermion}{v2,o2}
  \fmf{photon}{v1,i2}
  \fmf{photon}{o1,v2}
\fmfv{decor.shape=circle,decor.filled=1,decor.size=.01w,label=$q_1$,label.dist=.08w}{i1} 
\fmfv{decor.shape=circle,decor.filled=1,decor.size=.01w,label=$p_2$,label.dist=.08w}{i2} 
\fmfv{decor.shape=circle,decor.filled=1,decor.size=.01w,label=$q_2$,label.dist=.08w}{o1}
\fmfv{decor.shape=circle,decor.filled=1,decor.size=.01w,label=$p_1$,label.dist=.08w}{o2}
      \end{fmfgraph*}
      }
    }
%%%%%%%%%%%%%%%%%%%% BETHE-SALPETER DIAGRAMS %%%%%%%%%%%%%%%%
\newcommand{\freeppd}{\hspace{4mm}
%  \framebox{
   \parbox{11mm}{
      \begin{fmfgraph*}(7,10)
             \fmfset{arrow_len}{3.0mm}
             \fmfset{wiggly_len}{1.7mm}
             \fmfset{dot_size}{0.8thick}
             \fmfleft{i1,i2} \fmfright{o1,o2}
             \fmfleft{v1,v2}
             \fmfright{v3,v4} 
  \fmf{phantom}{v1,v2}
  \fmf{phantom}{v3,v4}
  \fmf{fermion}{v2,v1} % antiparticle to the left down 
  \fmf{fermion}{v3,v4} % particle to the right up
\fmfv{decor.shape=circle,decor.filled=1,decor.size=.01w,label=$q_1$,label.dist=.08w}{v1} 
\fmfv{decor.shape=circle,decor.filled=1,decor.size=.01w,label=$p_1$,label.dist=.08w}{v2} 
\fmfv{decor.shape=circle,decor.filled=1,decor.size=.01w,label=$q_2$,label.dist=.08w}{v3}
\fmfv{decor.shape=circle,decor.filled=1,decor.size=.01w,label=$p_2$,label.dist=.08w}{v4}
      \end{fmfgraph*}
      }
% }
    }
%%%%%%%%%%%%%%%%%%%%% version 2 %%%%%%%%%%%%%%%%%%%%%%%%%%%%%
\newcommand{\freeppdB}{\hspace{4mm}
   \parbox{11mm}{
      \begin{fmfgraph*}(7,10)
             \fmfset{arrow_len}{3.0mm}
             \fmfset{wiggly_len}{1.7mm}
             \fmfset{dot_size}{0.8thick}
             \fmfleft{i1,i2} \fmfright{o1,o2}
             \fmfleft{v1,v2}
             \fmfright{v3,v4} 
  \fmf{phantom}{v1,v2}
  \fmf{phantom}{v3,v4}
  \fmf{fermion}{v2,v1} % antiparticle to the left down 
  \fmf{fermion}{v3,v4} % particle to the right up
\fmfv{decor.shape=circle,decor.filled=1,decor.size=.01w,label=$q_1$,label.dist=.08w,label.angle=-90}{v1} 
\fmfv{decor.shape=circle,decor.filled=1,decor.size=.01w,label=$p_1$,label.dist=.08w,label.angle=90}{v2} 
\fmfv{decor.shape=circle,decor.filled=1,decor.size=.01w,label=$q_2$,label.dist=.08w,label.angle=-90}{v3}
\fmfv{decor.shape=circle,decor.filled=1,decor.size=.01w,label=$p_2$,label.dist=.08w,label.angle=90}{v4}
      \end{fmfgraph*}
      }
    }
%%%%%%%%%%%%%%%%%%%%%%%%%%%%%%%%%%%%%%%%%%%%%%%%%%%%%%%%%%%%%
\newcommand{\freePP}{\hspace{4mm}
   \parbox{11mm}{
      \begin{fmfgraph*}(7,10)
             \fmfset{arrow_len}{2.5mm}
             \fmfset{wiggly_len}{1.7mm}
             \fmfset{dot_size}{0.8thick}
             \fmfleft{i1,i2} \fmfright{o1,o2}
             \fmfleft{v1,v2}
             \fmfright{v3,v4} 
  \fmf{phantom}{v1,v2}
  \fmf{phantom}{v3,v4}
  \fmf{photon}{v1,v2}
  \fmf{photon}{v3,v4}
\fmfv{decor.shape=circle,decor.filled=1,decor.size=.01w,label=$q_1$,label.dist=.08w}{v1} 
\fmfv{decor.shape=circle,decor.filled=1,decor.size=.01w,label=$p_1$,label.dist=.08w}{v2} 
\fmfv{decor.shape=circle,decor.filled=1,decor.size=.01w,label=$q_2$,label.dist=.08w}{v3}
\fmfv{decor.shape=circle,decor.filled=1,decor.size=.01w,label=$p_2$,label.dist=.08w}{v4}
      \end{fmfgraph*}
      }
    }
%%%%%%%%%%%%%%%%%%%%%%%%%%%%%%%%%%%%%%%%%%%%%%%%%%%%%%%%%%%%%
\newcommand{\freePPB}{\hspace{4mm}
   \parbox{11mm}{
      \begin{fmfgraph*}(7,10)
             \fmfset{arrow_len}{2.5mm}
             \fmfset{wiggly_len}{1.7mm}
             \fmfset{dot_size}{0.8thick}
             \fmfleft{i1,i2} \fmfright{o1,o2}
             \fmfleft{v1,v2}
             \fmfright{v3,v4} 
  \fmf{phantom}{v1,v2}
  \fmf{phantom}{v3,v4}
  \fmf{photon}{v1,v2}
  \fmf{photon}{v3,v4}
\fmfv{decor.shape=circle,decor.filled=1,decor.size=.01w,label=$q_1$,label.dist=.08w,label.angle=-90}{v1} 
\fmfv{decor.shape=circle,decor.filled=1,decor.size=.01w,label=$p_1$,label.dist=.08w,label.angle=90}{v2} 
\fmfv{decor.shape=circle,decor.filled=1,decor.size=.01w,label=$q_2$,label.dist=.08w,label.angle=-90}{v3}
\fmfv{decor.shape=circle,decor.filled=1,decor.size=.01w,label=$p_2$,label.dist=.08w,label.angle=90}{v4}
      \end{fmfgraph*}
      }
    }
%%%%%%%%%%%%%%%%%%%%%%%%%%%%%%%%%%%%%%%%%%%%%%%%%%%%%%%%%%%%%
\newcommand{\crPP}{\hspace{4mm}
   \parbox{11mm}{
      \begin{fmfgraph*}(7,10)
             \fmfset{arrow_len}{2.5mm}
             \fmfset{wiggly_len}{1.7mm}
             \fmfset{dot_size}{0.8thick}
             \fmfleft{i1,i2} \fmfright{o1,o2}
             \fmfleft{v1,v2}
             \fmfright{v3,v4} 
  \fmf{phantom}{v1,v2}
  \fmf{phantom}{v3,v4}
  \fmf{photon}{v1,v4}
  \fmf{photon,rubout}{v3,v2}
\fmfv{decor.shape=circle,decor.filled=1,decor.size=.01w,label=$q_1$,label.dist=.08w}{v1} 
\fmfv{decor.shape=circle,decor.filled=1,decor.size=.01w,label=$p_1$,label.dist=.08w}{v2} 
\fmfv{decor.shape=circle,decor.filled=1,decor.size=.01w,label=$q_2$,label.dist=.08w}{v3}
\fmfv{decor.shape=circle,decor.filled=1,decor.size=.01w,label=$p_2$,label.dist=.08w}{v4}
      \end{fmfgraph*}
      }
    }
%%%%%%%%%%%%%%%%%%%%%%%%%%%%%%%%%%%%%%%%%%%%%%%%%%%%%%%%%%%%%
\newcommand{\crPPB}{\hspace{4mm}
   \parbox{11mm}{
      \begin{fmfgraph*}(7,10)
             \fmfset{arrow_len}{2.5mm}
             \fmfset{wiggly_len}{1.7mm}
             \fmfset{dot_size}{0.8thick}
             \fmfleft{i1,i2} \fmfright{o1,o2}
             \fmfleft{v1,v2}
             \fmfright{v3,v4} 
  \fmf{phantom}{v1,v2}
  \fmf{phantom}{v3,v4}
  \fmf{photon}{v1,v4}
  \fmf{photon,rubout}{v3,v2}
\fmfv{decor.shape=circle,decor.filled=1,decor.size=.01w,label=$q_1$,label.dist=.08w,label.angle=-90}{v1} 
\fmfv{decor.shape=circle,decor.filled=1,decor.size=.01w,label=$p_1$,label.dist=.08w,label.angle=90}{v2} 
\fmfv{decor.shape=circle,decor.filled=1,decor.size=.01w,label=$q_2$,label.dist=.08w,label.angle=-90}{v3}
\fmfv{decor.shape=circle,decor.filled=1,decor.size=.01w,label=$p_2$,label.dist=.08w,label.angle=90}{v4}
      \end{fmfgraph*}
      }
    }
%%%%%%%%%%%%%%%%%%%%%%%%%%%%%%%%%%%%%%%%%%%%%%%%%%%%%%%%%%%%%
\newcommand{\treeppd}{\hspace{4mm}
   \parbox{11mm}{
      \begin{fmfgraph*}(7,10)
             \fmfset{arrow_len}{2.5mm}
             \fmfset{wiggly_len}{1.7mm}
             \fmfset{dot_size}{0.8thick}
             \fmfleft{i1,i2} \fmfright{o1,o2}
  \fmf{fermion}{i2,v1,i1}
  \fmf{fermion}{o1,v2,o2}
  \fmf{photon}{v1,v2}
  \fmfdotn{v}{2}
\fmfv{decor.shape=circle,decor.filled=1,decor.size=.01w,label=$q_1$,label.dist=.08w}{i1} 
\fmfv{decor.shape=circle,decor.filled=1,decor.size=.01w,label=$p_1$,label.dist=.08w}{i2} 
\fmfv{decor.shape=circle,decor.filled=1,decor.size=.01w,label=$q_2$,label.dist=.08w}{o1}
\fmfv{decor.shape=circle,decor.filled=1,decor.size=.01w,label=$p_2$,label.dist=.08w}{o2}
      \end{fmfgraph*}
      }
    }
%%%%%%%%%%%%%%%%%%%%%%%%%%%%%%%%%%%%%%%%%%%%%%%%%%%%%%%%%%%%%
\newcommand{\treeppdB}{\hspace{4mm}
   \parbox{11mm}{
      \begin{fmfgraph*}(7,10)
             \fmfset{arrow_len}{2.5mm}
             \fmfset{wiggly_len}{1.7mm}
             \fmfset{dot_size}{0.8thick}
             \fmfleft{i1,i2} \fmfright{o1,o2}
  \fmf{fermion}{i2,v1,i1}
  \fmf{fermion}{o1,v2,o2}
  \fmf{photon}{v1,v2}
  \fmfdotn{v}{2}
\fmfv{decor.shape=circle,decor.filled=1,decor.size=.01w,label=$q_1$,label.dist=.08w,label.angle=-90}{i1} 
\fmfv{decor.shape=circle,decor.filled=1,decor.size=.01w,label=$p_1$,label.dist=.08w,label.angle=90}{i2} 
\fmfv{decor.shape=circle,decor.filled=1,decor.size=.01w,label=$q_2$,label.dist=.08w,label.angle=-90}{o1}
\fmfv{decor.shape=circle,decor.filled=1,decor.size=.01w,label=$p_2$,label.dist=.08w,label.angle=90}{o2}
      \end{fmfgraph*}
      }
    }
%%%%%%%%%%%%%%%%%%%%%%%%%%%%%%%%%%%%%%%%%%%%%%%%%%%%%%%%%%%%%
\newcommand{\treePPppd}{\hspace{4mm}
   \parbox{11mm}{
      \begin{fmfgraph*}(7,10)
             \fmfset{arrow_len}{2.5mm}
             \fmfset{wiggly_len}{1.7mm}
             \fmfset{dot_size}{0.8thick}
             \fmfleft{i1,i2} \fmfright{o1,o2}
  \fmf{fermion}{i2,v1,v2,o2}
  \fmf{photon}{i1,v1}  \fmf{photon}{o1,v2}
  \fmfdotn{v}{2}
\fmfv{decor.shape=circle,decor.filled=1,decor.size=.01w,label=$q_1$,label.dist=.08w}{i1} 
\fmfv{decor.shape=circle,decor.filled=1,decor.size=.01w,label=$p_1$,label.dist=.08w}{i2} 
\fmfv{decor.shape=circle,decor.filled=1,decor.size=.01w,label=$q_2$,label.dist=.08w}{o1}
\fmfv{decor.shape=circle,decor.filled=1,decor.size=.01w,label=$p_2$,label.dist=.08w}{o2}
      \end{fmfgraph*}
      }
    }
%%%%%%%%%%%%%%%%%%%%%%%%%%%%%%%%%%%%%%%%%%%%%%%%%%%%%%%%%%%%%
\newcommand{\treePPppdB}{\hspace{4mm}
   \parbox{11mm}{
      \begin{fmfgraph*}(7,10)
             \fmfset{arrow_len}{2.5mm}
             \fmfset{wiggly_len}{1.7mm}
             \fmfset{dot_size}{0.8thick}
             \fmfleft{i1,i2} \fmfright{o1,o2}
  \fmf{fermion}{i2,v1,v2,o2}
  \fmf{photon}{i1,v1}  \fmf{photon}{o1,v2}
  \fmfdotn{v}{2}
\fmfv{decor.shape=circle,decor.filled=1,decor.size=.01w,label=$q_1$,label.dist=.08w,label.angle=-90}{i1} 
\fmfv{decor.shape=circle,decor.filled=1,decor.size=.01w,label=$p_1$,label.dist=.08w,label.angle=90}{i2} 
\fmfv{decor.shape=circle,decor.filled=1,decor.size=.01w,label=$q_2$,label.dist=.08w,label.angle=-90}{o1}
\fmfv{decor.shape=circle,decor.filled=1,decor.size=.01w,label=$p_2$,label.dist=.08w,label.angle=90}{o2}
      \end{fmfgraph*}
      }
    }
%%%%%%%%%%%%%%%%%%%%%%%%%%%%%%%%%%%%%%%%%%%%%%%%%%%%%%%%%%%%%
\newcommand{\treePPcppd}{\hspace{4mm}
   \parbox{13mm}{
      \begin{fmfgraph*}(9,10)
             \fmfset{arrow_len}{2.5mm}
             \fmfset{wiggly_len}{1.7mm}
             \fmfset{dot_size}{0.8thick}
             \fmfleft{i1,i2} \fmfright{o1,o2}
  \fmf{fermion}{i2,v1}
  \fmf{fermion,tension=0.5}{v1,v2}
  \fmf{fermion}{v2,o2} 
  \fmf{phantom}{i1,v1}  \fmf{phantom}{o1,v2}
  \fmffreeze
  \fmf{photon}{i1,v2}  \fmf{photon,rubout}{o1,v1}
  \fmfdotn{v}{2}
\fmfv{decor.shape=circle,decor.filled=1,decor.size=.01w,label=$q_1$,label.dist=.08w}{i1} 
\fmfv{decor.shape=circle,decor.filled=1,decor.size=.01w,label=$p_1$,label.dist=.08w}{i2} 
\fmfv{decor.shape=circle,decor.filled=1,decor.size=.01w,label=$q_2$,label.dist=.08w}{o1}
\fmfv{decor.shape=circle,decor.filled=1,decor.size=.01w,label=$p_2$,label.dist=.08w}{o2}
      \end{fmfgraph*}
      }
    }
%%%%%%%%%%%%%%%%%%%%%%%%%%%%%%%%%%%%%%%%%%%%%%%%%%%%%%%%%%%%%
\newcommand{\treePPcppdB}{\hspace{4mm}
   \parbox{13mm}{
      \begin{fmfgraph*}(9,10)
             \fmfset{arrow_len}{2.5mm}
             \fmfset{wiggly_len}{1.7mm}
             \fmfset{dot_size}{0.8thick}
             \fmfleft{i1,i2} \fmfright{o1,o2}
  \fmf{fermion}{i2,v1}
  \fmf{fermion,tension=0.5}{v1,v2}
  \fmf{fermion}{v2,o2} 
  \fmf{phantom}{i1,v1}  \fmf{phantom}{o1,v2}
  \fmffreeze
  \fmf{photon}{i1,v2}  \fmf{photon,rubout}{o1,v1}
  \fmfdotn{v}{2}
\fmfv{decor.shape=circle,decor.filled=1,decor.size=.01w,label=$q_1$,label.dist=.08w,label.angle=-90}{i1} 
\fmfv{decor.shape=circle,decor.filled=1,decor.size=.01w,label=$p_1$,label.dist=.08w,label.angle=90}{i2} 
\fmfv{decor.shape=circle,decor.filled=1,decor.size=.01w,label=$q_2$,label.dist=.08w,label.angle=-90}{o1}
\fmfv{decor.shape=circle,decor.filled=1,decor.size=.01w,label=$p_2$,label.dist=.08w,label.angle=90}{o2}
      \end{fmfgraph*}
      }
    }
%%%%%%%%%%%%%%%%%%%%%%%%%%%%%%%%%%%%%%%%%%%%%%%%%%%%%%%%%%%%%
\newcommand{\treeppdPP}{\hspace{4mm}
   \parbox{11mm}{
      \begin{fmfgraph*}(7,10)
             \fmfset{arrow_len}{2.5mm}
             \fmfset{wiggly_len}{1.7mm}
             \fmfset{dot_size}{0.8thick}
             \fmfleft{i1,i2} \fmfright{o1,o2}
  \fmf{fermion}{o1,v2,v1,i1} 
  \fmf{photon}{v1,i2}  \fmf{photon}{v2,o2}
  \fmfdotn{v}{2}
\fmfv{decor.shape=circle,decor.filled=1,decor.size=.01w,label=$q_1$,label.dist=.08w}{i1} 
\fmfv{decor.shape=circle,decor.filled=1,decor.size=.01w,label=$p_1$,label.dist=.08w}{i2} 
\fmfv{decor.shape=circle,decor.filled=1,decor.size=.01w,label=$q_2$,label.dist=.08w}{o1}
\fmfv{decor.shape=circle,decor.filled=1,decor.size=.01w,label=$p_2$,label.dist=.08w}{o2}
      \end{fmfgraph*}
      }
    }
%%%%%%%%%%%%%%%%%%%%%%%%%%%%%%%%%%%%%%%%%%%%%%%%%%%%%%%%%%%%%
\newcommand{\treeppdPPB}{\hspace{4mm}
   \parbox{11mm}{
      \begin{fmfgraph*}(7,10)
             \fmfset{arrow_len}{2.5mm}
             \fmfset{wiggly_len}{1.7mm}
             \fmfset{dot_size}{0.8thick}
             \fmfleft{i1,i2} \fmfright{o1,o2}
  \fmf{fermion}{o1,v2,v1,i1} 
  \fmf{photon}{v1,i2}  \fmf{photon}{v2,o2}
  \fmfdotn{v}{2}
\fmfv{decor.shape=circle,decor.filled=1,decor.size=.01w,label=$q_1$,label.dist=.08w,label.angle=-90}{i1} 
\fmfv{decor.shape=circle,decor.filled=1,decor.size=.01w,label=$p_1$,label.dist=.08w,label.angle=90}{i2} 
\fmfv{decor.shape=circle,decor.filled=1,decor.size=.01w,label=$q_2$,label.dist=.08w,label.angle=-90}{o1}
\fmfv{decor.shape=circle,decor.filled=1,decor.size=.01w,label=$p_2$,label.dist=.08w,label.angle=90}{o2}
      \end{fmfgraph*}
      }
    }
%%%%%%%%%%%%%%%%%%%%%%%%%%%%%%%%%%%%%%%%%%%%%%%%%%%%%%%%%%%%%
\newcommand{\treeppdPPc}{\hspace{4mm}
   \parbox{13mm}{
      \begin{fmfgraph*}(9,10)
             \fmfset{arrow_len}{2.5mm}
             \fmfset{wiggly_len}{1.7mm}
             \fmfset{dot_size}{0.8thick}
             \fmfleft{i1,i2} \fmfright{o1,o2}
  \fmf{fermion}{o1,v2}
  \fmf{fermion,tension=0.5}{v2,v1}
  \fmf{fermion}{v1,i1} 
  \fmf{phantom}{i2,v1}  \fmf{phantom}{o2,v2}
  \fmffreeze
  \fmf{photon}{i2,v2}  \fmf{photon,rubout}{o2,v1}
  \fmfdotn{v}{2}
\fmfv{decor.shape=circle,decor.filled=1,decor.size=.01w,label=$q_1$,label.dist=.08w}{i1} 
\fmfv{decor.shape=circle,decor.filled=1,decor.size=.01w,label=$p_1$,label.dist=.08w}{i2} 
\fmfv{decor.shape=circle,decor.filled=1,decor.size=.01w,label=$q_2$,label.dist=.08w}{o1}
\fmfv{decor.shape=circle,decor.filled=1,decor.size=.01w,label=$p_2$,label.dist=.08w}{o2}
      \end{fmfgraph*}
      }
    }
%%%%%%%%%%%%%%%%%%%%%%%%%%%%%%%%%%%%%%%%%%%%%%%%%%%%%%%%%%%%%
\newcommand{\treeppdPPcB}{\hspace{4mm}
   \parbox{13mm}{
      \begin{fmfgraph*}(9,10)
             \fmfset{arrow_len}{2.5mm}
             \fmfset{wiggly_len}{1.7mm}
             \fmfset{dot_size}{0.8thick}
             \fmfleft{i1,i2} \fmfright{o1,o2}
  \fmf{fermion}{o1,v2}
  \fmf{fermion,tension=0.5}{v2,v1}
  \fmf{fermion}{v1,i1} 
  \fmf{phantom}{i2,v1}  \fmf{phantom}{o2,v2}
  \fmffreeze
  \fmf{photon}{i2,v2}  \fmf{photon,rubout}{o2,v1}
  \fmfdotn{v}{2}
\fmfv{decor.shape=circle,decor.filled=1,decor.size=.01w,label=$q_1$,label.dist=.08w,label.angle=-90}{i1} 
\fmfv{decor.shape=circle,decor.filled=1,decor.size=.01w,label=$p_1$,label.dist=.08w,label.angle=90}{i2} 
\fmfv{decor.shape=circle,decor.filled=1,decor.size=.01w,label=$q_2$,label.dist=.08w,label.angle=-90}{o1}
\fmfv{decor.shape=circle,decor.filled=1,decor.size=.01w,label=$p_2$,label.dist=.08w,label.angle=90}{o2}
      \end{fmfgraph*}
      }
    }
%%%%%%%%%%%%%%%%%%%%%%%%%%%%%%%%%%%%%%%%%%%%%%%%%%%%%%%%%%%%%%%%%%%%%%%%%%%%%%%
%%%%%%%%%%%%%%%%%%%%%%%%%%%%%%%%%%%%%%%%%%%%%%%%%%%%%%%%%%%%%%%%%%%%%%%%%%%%%%%

\begin{fmffile}{diags}

\renewcommand{\thesubsection}{\arabic{subsection}}
\newcommand{\be}{\begin{eqnarray}}
\newcommand{\ee}{\end{eqnarray}}
\newcommand{\ba}{\begin{array}}
\newcommand{\ea}{\end{array}}
\newcommand{\nn}{\nonumber}
\newcommand{\ra}{\rightarrow}
\newcommand{\lsim}{\;\parbox{0.35cm}{$\displaystyle\stackrel{\textstyle <}
{\sim}$}\;}
\newcommand{\gsim}{\;\parbox{0.35cm}{$\displaystyle\stackrel{\textstyle >}
{\sim}$}\;}

\def\s{\scriptstyle}
\def\ss{\scriptscriptstyle}
\def\ds{\displaystyle}
\def\p{\partial}

\begin{titlepage}
\setcounter{page}{0}
\thispagestyle{empty}

\begin{flushright}
November 11, 2000
\end{flushright}
\vspace{0.3in}

\begin{center}
\LARGE \bf Bound States from Regge Trajectories \\[1.5mm]
in a Scalar Model{\renewcommand{\thefootnote}
{\fnsymbol{footnote}}\footnotemark[2]\footnotetext[2]{This work was
supported by Conacyt grants 3298P-E9608 and 3298P-32729-E and also 
CIC-UMSNH.}}\renewcommand{\thefootnote}{\alph{footnote}} \\[0.25in]
\large \rm A. Weber\footnote{Part of the work done while at the Instituto de
Ciencias Nucleares, UNAM; e-mail: axel@io.ifm.umich.mx}\renewcommand
{\thefootnote}{,\fnsymbol{footnote}}\footnotemark[1]\renewcommand{\thefootnote}
{\alph{footnote}},
\large \rm J.C. L\'opez Vieyra\footnote
{e-mail: vieyra@pythia.nuclecu.unam.mx}\renewcommand{\thefootnote}
{,\fnsymbol{footnote}}\footnotemark[7]\renewcommand{\thefootnote}
{\alph{footnote}},
C.R.\ Stephens\footnote{e-mail: stephens@nuclecu.unam.mx}\renewcommand
{\thefootnote}{,\fnsymbol{footnote}}\footnotemark[7]\renewcommand{\thefootnote}
{\alph{footnote}},
S. Dilcher\footnote{Supported during part of the work
by fellowships of the DAAD and the Mexican
Government; present address:
Fakult\"at f\"ur Physik, A. Ludwig Universit\"at, %H.Herder Strasse 4,
79104 Freiburg, Germany; e-mail: dilcher@physik.uni-freiburg.de}\renewcommand
{\thefootnote}{,\fnsymbol{footnote}}\footnotemark[7]\renewcommand{\thefootnote}
{\alph{footnote}}, \\
P.O. Hess\footnote{e-mail: hess@nuclecu.unam.mx}\renewcommand
{\thefootnote}{,\fnsymbol{footnote}}\footnotemark[7] \\[0.15in]
\normalsize \it \renewcommand{\thefootnote}{\fnsymbol
{footnote}}\footnotemark[1]Instituto de F\'\i sica y Matem\'aticas, UMSNH,
58040 Morelia, Michoac\'an, Mexico, \\[0.5mm]
\footnotemark[7]\renewcommand{\thefootnote}{\arabic{footnote}}Instituto de
Ciencias Nucleares, UNAM, 04510 M\'exico D.F., Mexico
\end{center}
\vskip 0.6truein
%\begin{sloppypar}
{\bf Abstract.} The calculation of bound state properties using
renormalization group techniques to compute the corresponding Regge
trajectories is presented. In particular, we investigate the bound states in
different charge sectors of a scalar theory with interaction
$\phi^{\dag}\phi\chi$. The resulting bound state spectrum is surprisingly
rich. Where possible we compare and contrast with known results of the
Bethe-Salpeter equation in the ladder approximation and, in the
non-relativistic limit, with the corresponding Schr\"odinger equation.

%\vskip 0.35truein \noindent
%Keywords: bound states, Regge trajectories, Bethe-Salpeter equation
\vfill

\end{titlepage}
\setcounter{footnote}{0}

\subsection{Introduction}

The formation of bound states from more fundamental constituents is one
of the most important phenomena in physics spanning a vast spectrum of energy 
scales, ranging from high energies, associated with QCD bound states and 
resonances, to low energies in, for example, superconductivity. In describing
completely the crossover between bound states and their constituents one has
to address two fundamental problems: i) the formation of a single bound state
in terms of its more fundamental constituents, and ii) the many-body collective
effects associated with many bound states. The former has traditionally been
attacked by way of the Bethe-Salpeter equation \cite{BS} (for an early
review see ref.\ \cite{nakan}) and the latter by using
effective field theory, as in the Ginzburg-Landau formulation of
superconductivity for instance.

The problems with the Bethe-Salpeter equation are well known and various,
especially
when it comes to having controlled, systematic approximation schemes that
preserve all relevant symmetries. On the other hand, with effective field
theory one loses contact with the original fundamental degrees of freedom.
Naturally,
any technique which could give fresh insight into either of these two aspects
of the bound state problem would be more than welcome.

Given the non-perturbative nature of the underlying problem one might expect a
non-perturbative tool such as the renormalization group (RG) to be useful.
The fundamental problem here, in terms of a Wilsonian coarse graining picture,
is to formulate an RG that coarse grains bound states at low energies and
their fundamental constituents at high energies. Some progress has been made
in this direction \cite{ellwanger}. The fundamental problem however remains.
In the more restricted case of a non-relativistic system a non-relativistic
effective field theory description allied to an RG strategy that isolates
contributions from relativistic and non-relativistic momentum scales has
been used to calculate quantities such as the hyperfine structure of
atoms such as muonium or positronium \cite{lepage}.

In this paper we present a new and different approach to the description of
bound states based on RG techniques. Specifically, we exploit
the indirect access that Regge theory gives to the calculation of bound state
properties. Regge theory is unique in bringing together two extremely different
kinematical regions.
Given the asymmetric behaviour of the two-particle scattering amplitude in
the ``Regge limit'', $t \to \infty$, $s$ fixed (where $s$ and $t$ are the
Mandelstam variables), it predicts the energy spectrum of the
bound states in the theory.\footnote{A brief review of Regge theory, tailored
to our present purposes, has been given in ref.\ \cite{us}. A more complete
presentation may be found in ref.\ \cite{collins}.}

Given the difficulty in predicting the existence and properties of bound
states, if the theoretical determination of Regge trajectories from the
asymptotic behaviour of the scattering amplitudes should turn out to be more
accessible, we may be able to devise a more convenient approach to their
exploration. Indeed, in spite of the well known relation
between bound states and asymmetric scattering there has been little done
in terms of computing bound state properties in relativistic systems using
calculated Regge
trajectories. In this sense the present paper serves as a feasibility proof
that it is in fact possible to study bound state properties, both
qualitatively and quantitatively, via Regge trajectories. Of course,
it should be mentioned that Regge
trajectories also play a prominent r\^ole in the description of high energy
processes in QCD in the experimentally important kinematical region
$x \ll 1$ (for recent data see ref.\ \cite{hera}). In the
present contribution, however, we will concentrate exclusively on the
bound state aspects of the theory.

Regge trajectories were originally calculated in quantum field theory
by a direct summation of leading logs (in $t$) \cite{polk} and, again, using 
the Bethe-Salpeter equation \cite{lees}. There was a renewal of interest in 
Regge trajectories associated with the BKFL equation \cite{lipat}, which 
recently has been solved to next to leading order \cite{lipat2}.
In ref.\ \cite{us} we presented a new methodology based on RG
methods for doing field theory calculations in the Regge limit.
It has been sometimes thought that the
RG was incapable of accessing the Regge limit. As we have explained in some
detail in ref.\ \cite{us}, this is partially due to a somewhat restrictive
view of what the RG is and does. Indeed, a generalization in the
spirit of ``environmentally friendly''
renormalization \cite{environm} turned out to be necessary for that purpose.

One of the major disadvantages associated with the summation of leading logs
and the Bethe-Salpeter equation in comparison with RG methods
lies in the difficulty of incorporating higher-order corrections in a
systematic and consistent way. In very simple cases, such as $\phi^3$,
an analysis of the combinatorics of the $n$-loop diagrams is not so difficult.
However, as soon as one treats more complicated,
i.e.\ realistic, theories, the RG may be expected to be
superior to other methods. Even in the case of the simple looking
$\phi^{\dag}\phi\chi$ scalar theory
considered in ref.\ \cite{us}, a summation of the leading logs by
hand looks forbidding. Only {\em a posteriori} have we been able to
verify by a recursive argument that the direct summation leads to the
same result as the RG, and also the Bethe-Salpeter equation.

In this paper we will use our methods to calculate the bound state
spectrum in a scalar theory. We will compare our results to those of the 
Bethe-Salpeter equation where possible and discuss the relative merits of the 
different techniques.

\subsection{Wick-Cutkosky Model}

To illustrate the methodology we consider a theory consisting of a complex
scalar
field $\phi$ of mass $m$, to which we assign charge one,
and a real scalar field $ \chi $ of mass $\mu$, with interaction
$g \phi^{\dag} \phi \chi$. The reason we choose
this theory is that it is devoid of complications due to gauge fields and
spin yet exhibits a highly non-trivial and rich bound state
spectrum. In \cite{us} we investigated the asymptotic behaviour of the
two-particle scattering amplitude $A(s,t)$ in this theory, where
$s$ and $t$ are the standard Mandelstam variables.
In the limit of large $t$ we found different types of asymptotic behaviour,
corresponding to a double charged, a single charged and a neutral
state in the $s$-channel. Other processes are related by charge conjugation
symmetry.

In this section, we consider the simplest of these cases, namely the charge 
two ($Q=2$) sector corresponding to a double charged particle\footnote
{Strictly speaking, the exchanged object is a Reggeon. Only in the case where 
its angular momentum takes on a physical value can the object be identified 
with a physical particle, be it a bound state or a resonance.} exchanged in 
the $s$-channel. In ref.\ \cite{us} we found for the leading even-signature 
Regge trajectory to one-loop order (after RG improvement)
\be
\alpha_2^+(s) = g^2 K_{mm}(s) - 1 \:, \label{trajecttwo}
\ee
where the function $K$ is
\be
K_{m_1 m_2}(s)
 =\frac{1}{4\pi^2} \, \frac{
 \arctan  \sqrt{ \ds \frac{s - (m_1 - m_2)^2}{(m_1 + m_2)^2 - s}}  }
{\sqrt{\left( s - (m_1 - m_2)^2 \right)
\left( (m_1 + m_2)^2 - s \right) }} \:.
\ee
The leading log contributions resummed by the trajectory (\ref{trajecttwo})
come from ladder diagrams with $\phi$-particles on the sides and
$\chi$-particles on the rungs. The function $K$ corresponds to a
two-dimensional bubble diagram resulting from the contraction of the
``d-lines'' (rungs of the ladder) and the corresponding dimensional reduction
in the Regge limit.

Due to the Bose statistics of the $\phi$ field there is no odd-signature
trajectory. However, we can easily generalize this result to the case of
two different charged particles with masses $m_1$ and $m_2$. The leading
trajectories of even {\it and}\/ odd signature are then given (to one loop) by
\be
\alpha_{W}(s) = g^2 K_{m_1 m_2}(s) - 1 \:. \label{traject}
\ee
This result for the Regge trajectory has been known (in the case of equal
masses) for a long time \cite{polk,lees}. Physically, the (exact) trajectory
is expected to describe the bound state spectrum of the Wick-Cutkosky model
\cite{wick} via
\be
\alpha_{W}(M^2) = l \:, \label{spectrum}
\ee
according to Regge's theory. Here $M$ is the mass of the bound state with spin
$l$, $l = 0,1,2,\ldots$. Surprisingly enough, we were not able to find a
discussion of the bound state spectrum following from (\ref{spectrum})
anywhere in the literature.

Before discussing eq.\ (\ref{spectrum}) several remarks are in order. First
of all, we use the term ``Wick-Cutkosky model'' somewhat indiscriminately,
regardless of the mass of the exchanged $\chi$-particle, and refering in
principle to the solution of the full Bethe-Salpeter equation for the bound
state problem rather than some approximation to it. However, it is only in the
massless case and in the ladder approximation that the Bethe-Salpeter equation
has been analytically solved \cite{wick}, and this special case constitutes
the proper Wick-Cutkosky model. Both the ladder approximation to
the Bethe-Salpeter equation and the one-loop approximation to the Regge
trajectory are supposed to reproduce the spectrum properly only if the
coupling $g$ is not too large. Furthermore, it should be noted that eq.\
(\ref{spectrum}) is not expected to give the complete spectrum of the theory, 
but rather the lowest mass eigenvalue to every
spin $l$. Finally, there is no dependence on the mass of the exchanged
$\chi$-particle in the one-loop approximation to the Regge trajectory as a
result of taking the limit $t \to \infty$. On the other hand, the ladder 
approximation does display
such a dependence. In other words, the one-loop approximation to the Regge
trajectory does not distinguish between a Coulomb-like and a Yukawa-like
potential. This appears to be a serious restriction for the phenomenological
application of the Regge formalism and will be discussed in more detail in the
following. One should, however, bear in mind that for the most interesting
theories (namely gauge theories) most of the exchanged particles are massless.

Let us now solve eq.\ (\ref{spectrum}) for the bound state mass $M$.
To this end, it is convenient to introduce the dimensionless variables
\be
e^2 = \frac{g^2}{16 \pi m_1 m_2} \:, \label{effcoupl}
\ee
a combination that is commonly used as the effective coupling in the
Wick-Cutkosky model, and
\be
y = \frac{(m_1 + m_2)^2 - (m_1 - m_2)^2}{(m_1 + m_2)^2 - M^2} \:.
\ee
The physical meaning of the latter quantity will become apparent below.
Note that we here explicitly exclude the cases $m_1=0$ and $m_2=0$. In the
limit
where $m_1\rightarrow 0$ or $m_2\rightarrow 0$ a perturbative expansion of
the Regge trajectories will break down due to the presence of infrared
divergences.
In this limit a further renormalization of the Regge trajectory itself will
be necessary. We will not discuss this issue further here.

Eq.\ (\ref{spectrum}) now takes the form
\be
\frac{e^2}{\pi} \, y \, \frac{\arctan \sqrt{y-1}}{\sqrt{y-1}} = l + 1 \:.
\label{dimless}
\ee
Solving eq.\ (\ref{dimless}) we compare, over the entire range $M=0$ to 
$M=m_1+m_2$, with known results from the ladder approximation to the
Bethe-Salpeter equation for the case $\mu=0$, $m_1=m_2=m$, and $l=0$ (the
ground state) \cite{linden}. The comparison
can be seen in fig.\ 1 where we also plot the Bethe-Salpeter results for
$\mu=m/10$. 
\begin{figure}
\begin{center}  %%% figure 1
\begin{picture}(90,70)
%\put(0,0){\framebox(90,70){}}
\put(3,-20){\psfig{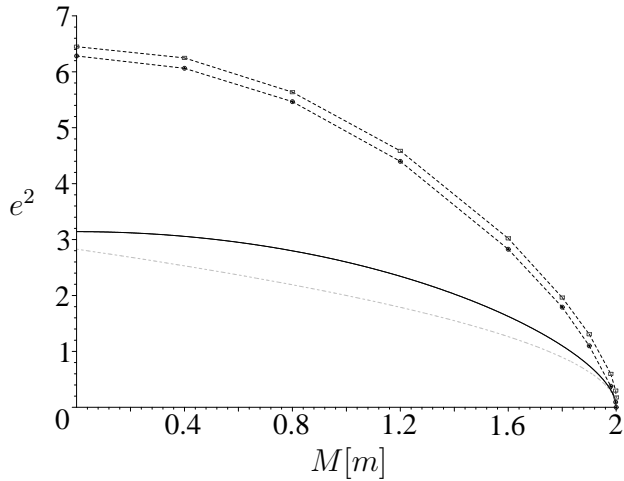}}
\put(43,3){$M [m]$}
\put(3,37){$e^2$}
\end{picture}
\end{center}
\vspace{-6mm}
\caption{Graph of coupling versus mass of the ground state for 
$m_1 = m_2 = m$. The solid line is from equation (\ref{dimless}) and the
dotted line from the non-relativistic Coulomb spectrum (with $M = m_1 + m_2
+ E$). The dashed curves represent the (numerical) solutions \cite{linden} of 
the Bethe-Salpeter equation, with $\mu = 0$ for the lower dashed curve and 
$\mu = m/10$ for the upper one.}
\end{figure}
There are two noteworthy features: first of all the
difference between our result and that of ref.\ \cite{linden}, although
relatively large, cannot be used to favour one methodology versus the other as 
the difference appears in the region of rather strong coupling 
($e^2 \gsim 0.5$, see also fig.\ 2 below). In this strong-coupling region
higher loop corrections will be important. We believe that
such corrections are easier to compute in the present RG approach than in
the Bethe-Salpeter equation. The second feature is that, as one can see from
comparing the two Bethe-Salpeter curves, the effect of $\mu\neq 0$ is
relatively weak, which leads us
to believe that our methodology, based on the Regge limit, should not
suffer too much from the fact that, at least at lowest order, our results
are $\mu$-independent. The $\mu$-dependence is, however, significant near
the threshold $M = 2m$.

We can gain a qualitative understanding of the situation from a simple
phenomenological consideration: a rough estimate for the extension of
the bound state is given by the Bohr radius
\be
a_l = \frac{l+1}{m_{rd} \, e^2} \label{bohr}
\ee
for states with principal quantum number $n=l+1$. Here $m_{rd}$ is the
usual reduced mass
\be
m_{rd} = \frac{m_1 m_2}{m_1 + m_2} \:. \label{effreduced}
\ee
Intuitively, for a state to remain bound, its radius cannot be larger than the 
range of the potential, hence $a_l \lsim R = 1/\mu$, or
\be
e^2 \gsim \frac{\mu}{m_{rd}} (l+1) \:. \label{existence}
\ee
In the case of a Yukawa-like potential ($\mu \neq 0$) this yields an
estimate of the number of existing bound states (with $n = l+1$), while
for $\mu = 0$ there is obviously no restriction.

Similarly, the finiteness of $\mu$ or the range of the potential is not
expected to be significant for the spectrum as long as $a_l \ll R$, or
equivalently
\be
e^2 \gg \frac{\mu}{m_{rd}} (l+1) \:. \label{estimate}
\ee
We should expect our $\mu$-independent results to be quantitatively correct
in this region (except for corrections from higher loop orders). In fact,
this expectation is borne out by fig.\ 1, where (\ref{estimate}) becomes
$e^2 \gg 0.2$ for $\mu = m/10$, and in this region the deviation from the
$\mu = 0$ curve is indeed small.

In the following, we will often consider the non-relativistic limit, simply
because it is relatively easy to obtain exact solutions in this limit. A system
will be considered non-relativistic whenever $p \ll m_{rd}$, where we can
obtain an estimate for the momentum from Heisenberg's uncertainty
relation, $p \approx 1/a_l$. With (\ref{bohr}), the condition for a system
to be non-relativistic becomes
\be
e^2 \ll l+1 \:.
\ee
Thus from (\ref{existence}) there can be non-relativistic bound states
only if
\be
\mu \ll m_{rd} \:. \label{nonrelcond}
\ee

Let us now consider the solutions of (\ref{spectrum}) in the weak-coupling 
or non-relativistic (for all $l$) limit $e^2 \ll 1$. Eq.\ (\ref{dimless})
implies that in this case the solutions satisfy $y \gg 1$. Consequently,
\be
\frac{(m_1 + m_2)^2 - M^2}{2(m_1 + m_2)} \ll 2 m_{rd} < m_1 + m_2 \:,
\label{energy}
\ee
where $m_{rd}$ is the reduced mass defined in (\ref{effreduced}). We can 
then identify the left-hand side of
(\ref{energy}) with the absolute value of the (negative) binding energy $E$,
considering that this definition is equivalent to
\be
M = m_1 + m_2 + E
\ee
in the limit $|E| \ll m_1 + m_2$.

Replacing $y$ by $2m_{rd}/|E|$ and considering the limit $y \gg 1$, the
solution to (\ref{dimless}) is easily found to be
\be
E = - \frac{m_{rd} \, e^4}{2 (l+1)^2} \:, \label{coulomb}
\ee
the well-known non-relativistic Coulomb spectrum (with principal quantum
number $n = l+1$).

The Bethe-Salpeter equation for a massless $\chi$ particle in the ladder
approximation leads, of course, to the same result in the limit $e^2\to 0$
\cite{wick}. For massive $\chi$, however, the threshold behavior is
qualitatively different: there exists a critical coupling $e_c^2$ below
which there are no bound states. In the limit $\mu \to 0$, one has
$e_c^2 \cong 0.84 (\mu/m_{rd})$ \cite{SZ} which is in excellent agreement
with the estimate (\ref{existence}). The existence of a critical coupling
is not accounted for by the one-loop formula for the Regge trajectory.
However, as soon as $e^2 \gg (\mu/m_{rd})$, the variation of the ground
state energy with $\mu$ will be quantitatively small.

In fig.\ 2 we compare the solution of eq.\ (\ref{dimless}) in the
non-relativistic region with  eq.\ (\ref{coulomb}) and once again the results
of \cite{linden} using the ladder approximation for $m_1=m_2$ and 
$m_1=100m_2$, and $\mu = 0$.
\begin{figure}
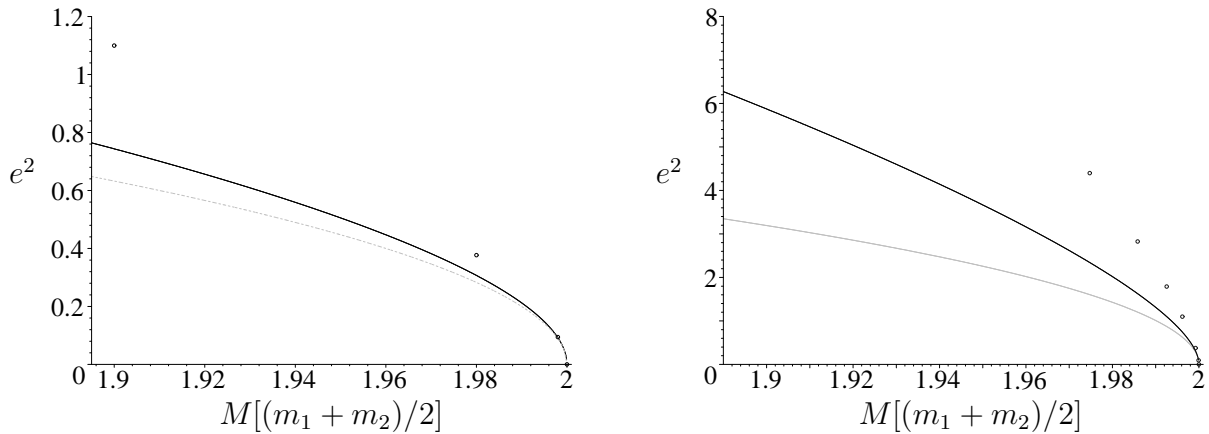
  %%% figure 2
\begin{center}
\begin{picture}(165,61)
%\put(0,0){\framebox(165,61){}}
%%% delta=1
%\put(0,0){\framebox(80,61){}}
\put(4,-17){\psfig{figure=FIGURE2a.ps,height=80mm,angle=-90}}
\put(29,3){$M [(m_1 + m_2)/2]$}
\put(1,35){$e^2$}
%%%  delta = 0.01
%\put(85,0){\framebox(80,61){}}
\put(88,-17){\psfig{figure=FIGURE2b.ps,height=80mm,angle=-90}}
\put(114,3){$M [(m_1 + m_2)/2]$}
\put(87,35){$e^2$}
\end{picture}
\end{center}
\vspace{-6mm}
\caption{Graph of coupling versus mass of the ground state for $m_1=m_2$, 
$\mu=0$ (left) and $m_1=100 \, m_2$, $\mu=0$ (right), respectively. The solid 
lines are from equation (7); the dashed curves are from the non-relativistic
Coulomb spectrum and the points from (numerical) solutions \cite{linden} of 
the Bethe-Salpeter equation.}
\end{figure}
Since we have not found any remark in the literature, let us note that the
numerical values for the coupling constants near threshold given in ref.\ 
\cite{linden} for the unequal mass case are in error (or the error in the
numerical computations is much larger than stated).\footnote{We thank Norbert
E. Ligterink for discussions on the asymmetric mass limit in the
Bethe-Salpeter equation.} We have produced the
data presented in fig.\ 2 by applying the Wick-Cutkosky transformation
\cite{nakan,wick} to the equal mass results given in ref.\ \cite{linden}.
Obviously, the Regge data lie in between the Bethe-Salpeter and the
non-relativistic curves. This is a welcome result in so far as virtually all
alternative approaches to relativistic bound state calculations yield values
in between these two curves.

\subsection{Charge One Bound States}

In the case of the exchange of a Reggeon with charge one in the $s$-channel,
we found in ref.\ \cite{us} the following trajectories
\be
\alpha_1^{\pm}(s)= \pm g^2 K_{m\mu}(s) - 1 \:. \label{reggeone}
\ee
The positive-signature trajectory $\alpha_1^+$ corresponds to an attractive
potential, while the neg\-a\-tive-signature trajectory $\alpha_1^-$ indicates
a repulsive interaction. Consequently, $\alpha_1^+$ leads to a series of
bound states, while $\alpha_1^-$ always has a negative real part (for real
values of $s$).

The appearance of a repulsive interaction (through exchange of spinless bosons)
is a known feature of exchange interactions. The physics of an exchange
interaction is clearly illustrated by the diagrams that contribute to leading
order to the two trajectories and which have the following typical appearance
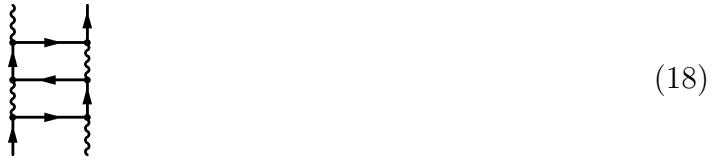
\be
\choneladder
\label{chargeone}
\ee
The bound state spectrum generated by the trajectory $\alpha_1^+$ is identical 
to the one obtained from $\alpha_W$ in the previous section, with $m_1$ and 
$m_2$ replaced by $m$ and $\mu$, respectively, in the definition
of the effective coupling and the reduced mass, except for the fact that $l$
can take only even values. This is the spectrum 
one would expect for $\phi\chi$ bound
states, with an interaction caused by the exchange of a massless particle.
But, as we know from the discussion of the Wick-Cutkosky model in the last
section, the Regge trajectory does not ``see'' the mass of the exchanged
particle, so that the interaction might as well originate from the exchange of
a $\phi$-particle with mass $m$, as one would have expected from the diagram
(\ref{chargeone}). 

The quantity $(\mu/m_{rd})$ that played a crucial r\^ole for the estimates
(\ref{existence})--(\ref{nonrelcond}), has to be replaced by $1 + (m/\mu)$ 
in the present situation. Consequently, no bound states 
are expected to exist for couplings
$e^2 \lsim 1 + (m/\mu)$, and on account of the missing dependence of the
Regge trajectory on the mass of the exchanged particle, our results cannot be 
expected to be accurate unless $e^2 \gg  1 + (m/\mu)$ (for $l=0$), in which 
case higher loop corrections will probably be important for 
quantitatively correct results. Thus it might appear that the results 
presented here for the charge one sector are of little use.
However, the nature of the present study is exploratory, and it is
legitimate to inquire whether other qualitative and quantitative
features are correctly
reproduced in the Regge approach. It will turn out that this is indeed the
case and suggests that one may consider the spectrum following from 
(\ref{reggeone}) as a valid first approximation to the real spectrum,
at least for $e^2 \gsim 1 + (m/\mu)$.

We will compare our results for the bound state masses in the different 
charge sectors in section 5. Here, in order to further analyse the physical
situation and the properties of the charge one bound states, 
we will formally consider the non-relativistic limit. In particular, 
the relation between the Regge signatures and the attractive 
or repulsive nature of the associated potentials will become obvious. We will 
start from the Bethe-Salpeter equation, which is easily written down in the 
ladder approximation for the present situation. In Euclidean momentum space, 
we have
\be
(p_1^2 + m^2) (p_2^2 + \mu^2) \Phi(p_1,p_2) = \int \frac{d^4 q_1}{(2\pi)^4}
\frac{d^4 q_2}{(2\pi)^4} \, V(p_1,p_2;q_1,q_2) \Phi(q_1,q_2) \:,
\label{BSchargeone}
\ee
where the Bethe-Salpeter kernel is given in this approximation by
\vspace{3mm}
\be
V(p_1,p_2;q_1,q_2) = \chonetree
= \frac{g^2}{(p_2 - q_1)^2 + m^2} \, (2\pi)^4 \delta(p_1 + p_2 - q_1 - q_2) 
\:. \label{kernel}
\ee
\vspace{1mm}

\noindent 
Reintroducing the speed of light $c$ and taking the limit $c\to \infty$ 
\cite{nakan} leads (after a Wick rotation to Minkowski space) 
to the Schr\"odinger equation
\be
-\frac{1}{2 m_{rd}} \nabla^2 \psi({\bf r}) - e^2 \, \frac{e^{-m r}}{r} \,
\psi(-{\bf r}) = E \psi({\bf r})
\ee
in the center-of-mass frame. After separation of the angular part, one
is left with the Schr\"odinger equation for the radial part,
\be
\left( -\frac{1}{2 m_{rd}} \frac{d^2}{d r^2} + \frac{l (l+1)}
{2 m_{rd} \, r^2} - (-1)^l e^2 \, \frac{e^{-m r}}{r} \right) R(r) = E R(r) 
\:, \label{exchschroed}
\ee
for angular momentum $l$. The presence of the factor $(-1)^l$ can be traced
directly to the exchange form of the kernel (\ref{kernel}). It is then clear
that we get a repulsive potential for $l$ odd, corresponding to the odd
signature trajectory, and an attractive one for $l$ even, corresponding to 
even signature, in accordance with the prediction of Regge theory. No 
solutions of eq.\ (\ref{BSchargeone}) are expected to exist 
in the non-relativistic regime.
If, however, one puts $m$ equal to zero by hand in the potential term (which
means neglecting the mass of the exchanged particle), the spectrum following
from (\ref{exchschroed}) is identical to the one obtained from 
$\alpha_1^+ (s)$ in the limit $e^2 \ll 1$.

To summarize, the Regge approach captures the qualitative and 
quantitative features of the physical situation, with the exception of those
related to the mass of the exchanged particle.

\subsection{Charge Zero Bound States: Equal Masses}

The exchange of charge zero Reggeons turned out to lead to a very
interesting set of leading Regge trajectories in ref.\ \cite{us}. We found
two trajectories of positive signature with a complicated analytical structure,
\be
\alpha^+_{0,\pm}(s) =
\frac{g^2 K_{mm}(s)}{2}\left(1 \pm \left( 1+8\frac{K_{\mu\mu}(s)}{K_{mm}(s)}
\right)^{\frac{1}{2}} \right) -1 \:, \label{possign}
\ee
and one trajectory of negative signature,
\be
\alpha_0^-(s) =g^2 K_{mm}(s)-1 \:.
\ee
The latter trajectory is identical in its functional form to $\alpha_2^+$,
whereas the signatures are different. This can be easily understood by
considering the contributing ladder diagrams
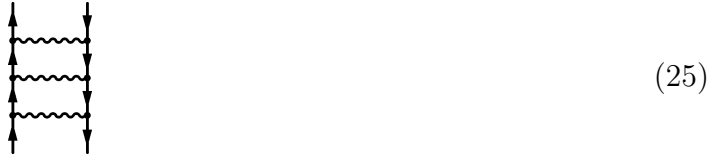
\be
\chzeroneg
\label{chargezeroodd}
\ee
in the non-relativistic limit. There can be no contribution with two $\chi$
particles on the sides of the ladder because their Bose statistics is
incompatible with the negative signature (odd angular momentum) of the
trajectory. The remaining diagrams (\ref{chargezeroodd}) then lead to the
same asymptotic behaviour as the ones for charge-two Reggeons, whilst the
latter can only have positive signature (even angular momentum) for the same
reason.

The positive-signature sector has a more interesting structure, and we were
in fact at first surprised to obtain two different trajectories in this case,
one corresponding to an attractive and the other to a repulsive potential.
Again, however, this qualitative feature can be better understood by
considering the non-relativistic limit of the corresponding Bethe-Salpeter
equation.

The typical ladder diagrams contributing to the Regge trajectory
(\ref{possign})
are of the form
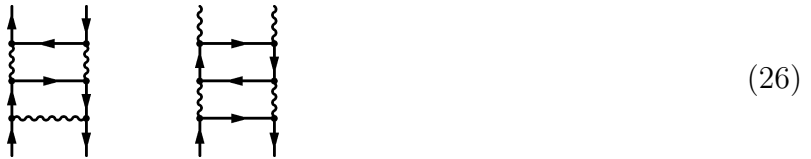
\be
\chzeroposA \hspace{1cm} \chzeroposB \label{chargezeroeven}
\ee
The Bethe-Salpeter equation that reproduces these diagrams most naturally has
a matrix structure to take care of the possible conversion of $\phi^\dagger
\phi$-states to $\chi\chi$-states. Such a matrix, or ``multi-channel'', 
Bethe-Salpeter equation has been used before
in ref.\ \cite{kaufm} for a (fictitious) coupled channel analysis and more
recently
in ref.\ \cite{kret}. It is also implicit in our earlier results 
\cite{us} where it was equivalent to implementing a matrix renormalization.

We will write the Bethe-Salpeter wave function as
\be
\Phi(p_1,p_2) =
\left( \begin{array}{c}
\Phi_1(p_1,p_2) \\
\Phi_2(p_1,p_2)
\end{array} \right) \:,
\ee
where the upper entry refers to the $\phi^\dagger \phi$-component and the
lower one to the $\chi\chi$-component of the wave function. In particular,
$\Phi_2$ has to be symmetric under interchange of its arguments because of
the Bose statistics for the $\chi$ particles.

In order that the ladder approximation to the Bethe-Salpeter equation reproduce
the diagrams (\ref{chargezeroeven}), we take for the two-particle propagator
(in Euclidean momentum space)
\be
G(p_1,p_2;q_1,q_2) =
\left(
\begin{array}{cc}
\rule{0mm}{9mm} \ \freeppd \ & \mbox{\Large 0} \\[30pt]
\mbox{\Large 0} & \mbox{\Large $\frac{1}{2}$} 
\left( \rule{0mm}{9mm} \ \freePP  +  \crPP \ \right)
\end{array}
\right)
\ee
and for the Bethe-Salpeter kernel to lowest order
\be
V(p_1,p_2;q_1,q_2) =
\left(
\begin{array}{cc}
\treeppd & \mbox{\Large $\frac{1}{2}$} 
\left(\rule{0mm}{9mm} \ \treePPppd + \treePPcppd \ \right) \\[30pt]
\rule[-7mm]{0mm}{17mm} \ \treeppdPP  +  \treeppdPPc \ &
\mbox{\Large 0}
\end{array}
\right)
\ee
The additional factor of $1/2$ in the kernel is due to phase space reduction
as a result of the identity of the $\chi$ particles and could equally have
been put in front of the other off-diagonal element of $V$, or as an additional
factor in front of the $\chi\chi$-propagator in $G$. In the present form the
non-relativistic limit turns out to have a particularly transparent
interpretation.

The Bethe-Salpeter equation in the ladder approximation now takes the 
following form,
\be
\lefteqn{(p_1^2 + m^2) (p_2^2 + m^2) \Phi_1(p_1, p_2)} \hspace{10mm} \nn \\
&=& g^2 \int \frac{d^4 q_1}{(2\pi)^4} \frac{1}{(p_1 - q_1)^2 + \mu^2} \,
\Phi_1(q_1, p_1 + p_2 - q_1) \nn \\
&& {}+ \frac{g^2}{2} \int \frac{d^4 q_1}{(2\pi)^4} \left( \frac{1}
{(p_1 - q_1)^2 + m^2} + \frac{1}{(p_2 - q_1)^2 + m^2} \right)
\Phi_2(q_1, p_1 + p_2 - q_1) \:, \nn \\[2mm]
\lefteqn{\frac{1}{2} \, (p_1^2 + \mu^2) (p_2^2 + \mu^2) \left[
\Phi_2(p_1, p_2) + \Phi_2(p_2, p_1) \right]} \hspace{10mm} \nn \\
&=& g^2 \int \frac{d^4 q_1}{(2\pi)^4} \left(
\frac{1}{(p_1 - q_1)^2 + m^2} + \frac{1}{(p_2 - q_1)^2 + m^2} \right)
\Phi_1(q_1, p_1 + p_2 - q_1) \:. \label{BSzero}
\ee
As in the previous section, for the purposes of a further
understanding of the physical situation
we will consider these equations in the non-relativistic limit,
in their Wick-rotated form in Minkowski space. Due to the possible
conversions between $\phi^\dagger\phi$-states and $\chi\chi$-states one
requires
that $\mu = m$
in order to have a consistent non-relativistic limit for both kinds of
states.\footnote{\label{nonrelmix}Actually, the condition
$|\mu - m| \ll m$ would be sufficient for this purpose. We will comment on
this case below.} However, we will continue to use different notations for
the two masses in the following.

In the formal limit $c \to \infty$, eqs.\ (\ref{BSzero}) lead to the following
coupled Schr\"odinger equations, written in the center-of-mass frame,
\be
-\frac{1}{m} \nabla^2 \psi_1({\bf r}) - \frac{g^2}{16 \pi m^2} 
\left( \frac{e^{-\mu r}}{r} \, \psi_1({\bf r}) + \frac{1}{2} 
\frac{e^{-mr}}{r} \left[ \psi_2({\bf r}) +
\psi_2(-{\bf r}) \right] \right) &=& E \, \psi_1({\bf r}) \:, \nn \\
-\frac{1}{2\mu} \nabla^2 \left[ \psi_2({\bf r}) + \psi_2(-{\bf r}) \right]
- \frac{g^2}{16 \pi \mu^2} \frac{e^{-mr}}{r} \left[ \psi_1({\bf r}) +
\psi_1(-{\bf r}) \right] &=& \frac{E}{2} \left[ \psi_2({\bf r}) +
\psi_2(-{\bf r}) \right] \nn \\
& & \label{schroedzero}
\ee
(the reduced masses are given by $m/2 $ and $\mu/2$ in this case). 
Note that only the
even component of the wave function $\psi_2$ appears, corresponding to the
part of $\Phi_2$ which is symmetric under exchange of the $\chi$ particles.
Consequently, the system (\ref{schroedzero}) has very different characteristics
for even and odd angular momentum.

To begin with the simpler case: for odd angular momentum the second equation
in (\ref{schroedzero}) becomes trivial and the contribution of $\psi_2$ to
the first equation cancels, so that one is left with the Schr\"odinger
equation for a Yukawa potential,
\be
-\frac{1}{2 m_{rd}} \nabla^2 \psi_1 ({\bf r}) - e^2 \, \frac{e^{-\mu r}}{r} \,
\psi_1 ({\bf r}) = E \psi_1 ({\bf r}) \:,
\ee
the definitions of $e^2$ and $m_{rd}$ being as in eqs.\ (\ref{effcoupl}) and
(\ref{effreduced}), with $m_1 = m_2 = m$. Hence the sitation is exactly as in
section 2 for the Wick-Cutkosky model, which is what we had concluded
before from a consideration of the diagrams alone. Let us remark that, in
analogy with the case of para-positronium, a decay into (or mixing with)
states of three $\chi$-particles is possible, but needs the consideration
of higher loop orders in the Regge trajectory.

For even angular momentum, and after separation of the angular part of the
wave functions, the radial Schr\"odinger equation takes the form
\be
\left[ -\frac{1}{2 m_{rd}} \frac{d^2}{d r^2} + \frac{l (l+1)}{2 m_{rd} r^2}
- e^2 \, \frac{e^{- \mu r}}{r} \left( \begin{array}{cc}
1 & 1 \\
2 & 0
\end{array} \right) \right] \left( \begin{array}{c}
R_1(r) \\ R_2(r)
\end{array} \right) = E \left( \begin{array}{c}
R_1(r) \\ R_2(r)
\end{array} \right) \:, \label{mixschroed}
\ee
where $e^2$ and $m_{rd}$ are defined as before, and we have now explicitly put
$\mu = m$. The matrix giving the potential is diagonalizable and has 
eigenvalues
\be
V_+(r) = -2 e^2 \, \frac{e^{- \mu r}}{r}  \:, \quad 
V_-(r) = +e^2 \, \frac{e^{- \mu r}}{r} \:. \label{diagpot}
\ee
Consequently, for one mixture of $\phi^\dagger\phi$- and $\chi\chi$-states
one has an attractive potential with bigger effective coupling constant
while for the other mixture the effective potential is repulsive. The
estimates of eqs.\ (\ref{existence})--(\ref{nonrelcond}) applied to the 
present case imply
that there will be no bound states for $e^2 \lsim 1$, in particular
eq.\ (\ref{BSzero}) is not expected to have solutions in the
non-relativistic regime.

We now compare these results with the Regge trajectories
(\ref{possign}), which for $\mu = m$ become
\be
\alpha^+_{0,+}(s) = 2 g^2 K_{mm}(s) - 1 \:, \quad
\alpha^+_{0,-}(s) = -g^2 K_{mm}(s) - 1 \:.
\ee
The first of these leads to a Coulomb spectrum with $e^2$ replaced by
$2 e^2$ in (\ref{coulomb}), while the second trajectory is
identical with $\alpha_1^-$ in (\ref{reggeone}) for $\mu = m$ and
corresponds to a repulsive potential. Hence, if we simply neglect the
mass in the exponential, the spectrum of bound states coincides with the 
spectrum generated by $V_+$ in (\ref{diagpot}). Furthermore, the 
Schr\"odinger equation (\ref{exchschroed}), which has been shown to correspond 
to $\alpha_1^-$, is the same as the one for the potential $V_-$.

Of course, we are not entitled to neglect the mass of the exchanged particle
unless $e^2 \gg 1$, and then higher order corrections give potentially
important contributions. However, one again arrives at the conclusion that
the Regge trajectories describe the physical situation correctly, except
for the effect of the mass of the exchanged particle.

An interesting side issue is related to the fact that the Hamiltonian in
(\ref{mixschroed}) is apparently non-hermitian.\footnote{We thank Jaime
Besprosvany for drawing our attention to this point.} The natural scalar
product of two non-relativistic two-component wave functions
$\psi({\bf r}_1, {\bf r}_2)$ and $\psi'({\bf r}_1, {\bf r}_2)$ depending on
the positions of the respective two particles, is given by
\be
& & \int d^3 r_1 \, d^3 r_2 \left( \psi_1^{\ast}({\bf r}_1, {\bf r}_2)
\psi_1'({\bf r}_1, {\bf r}_2) + \frac{1}{2} \psi_2^{\ast}({\bf r}_1, {\bf r}_2)
\psi_2'({\bf r}_1, {\bf r}_2) \right) \nn \\
&=& \int d^3 r_1 \, d^3 r_2 \, \psi^{\dag}({\bf r}_1, {\bf r}_2) \left(
\begin{array}{cc}
1 & 0 \\
0 & 1/2
\end{array} \right) \psi'({\bf r}_1, {\bf r}_2) \:,
\ee
due to the identity of the $\chi$ particles. For this scalar product, the
condition of hermiticity for an operator $H$ reads
\be
\left( \begin{array}{cc}
1 & 0 \\
0 & 1/2
\end{array} \right) H = H^{\dag} \left( \begin{array}{cc}
1 & 0 \\
0 & 1/2
\end{array} \right) \:,
\ee
and one can easily check that the Hamiltonian in (\ref{mixschroed}) is indeed
hermitian in this sense.

\subsection{Charge Zero Bound States: Unequal Masses}

In this section, we will comment on the case of different masses $\mu \neq m$
in the charge zero sector. As shown in ref.\ \cite{us}, in
this case there are two thresholds at $s = 4 \mu^2$ and $4 m^2$,
and the trajectory $\alpha_{0,+}^+$ leads to a series of bound states (below
both thresholds) and a series of resonances (between the thresholds).

An analysis of the equation
\be
\alpha_{0,+}^+ (M^2) = l \label{chzerogeneral}
\ee
in the weak-coupling limit
\be
e^2 = \frac{g^2}{16 \pi m^2} \to 0
\ee
in analogy to the analysis of eq.\ (\ref{spectrum}) in section 2, leads to
possible solutions in the two cases ($E < 0$)
\be
M = 2m + E \:, \quad |E| \ll m \label{mthresh}
\ee
and
\be
M = 2\mu + E \:, \quad |E| \ll \mu \:. \label{muthresh}
\ee

In the first case, i.e.\ near the threshold $4m^2$, the spectrum reduces to
the Coulomb spectrum (\ref{coulomb}), with $m_{rd} = m/2$, as long as
$e^4 \ll |m - \mu|/m$. There is a correction term of higher order in $e^2$,
which is real for $m < \mu$ and complex for $m > \mu$. This complex value
of $E$ indicates the instability of the corresponding resonance state.
The sign of the imaginary part is in agreement with expectations, whereas
for its size one would have expected a higher power of the effective coupling
constant from semi-phenomenological considerations.\footnote{By
``semi-phenomenological considerations'' we refer to the calculation of the
imaginary part as it has been performed for example for positronium, 
using Fermi's Golden Rule. In this case
one argues that the decay amplitude is proportional to the square of the
overlap between the (non-relativistic) wave functions
of the positron and electron. This overlap factor leads to a high power
of the coupling constant in the decay rate.}
Despite this descrepency, it is a highly attractive feature of the Regge
methodology that it describes bound states and resonances on the same footing,
thereby (in principle) furnishing information on the decay of the latter, and
may play an important r\^ole in future applications.

Somewhat exotic, or maybe pathological, features arise when one considers
the limit $m^2 \gg \mu^2$ (still in the case (\ref{mthresh})). The spectrum,
which remains Coulombic as long as $e^2 \ln (m^2/\mu^2) \ll 1$, takes the
form
\be
E = \frac{m_{rd} e^4}{2 (l+1)^2} \left( 1 - \frac{e^2}{\pi (l+1)} \ln
\frac{m^2}{\mu^2} \right)^2 \:,
\ee
provided that
\be
\frac{2 e^2}{\pi} \ln \frac{m^2}{\mu^2} \le l + 1 \:.
\ee
For larger values of $e^2 \ln (m^2/\mu^2)$, the solutions have a large
imaginary part. Hence in this somewhat artificial limit the situation may
arise that a series of resonances exist with angular momenta above a certain
minimal value, while states corresponding to lower angular momenta are
too unstable to be interpretable as propagating (and decaying) particles.
At the present stage of the investigation, it is impossible to tell if there
is any situation in the real world where this kind of behaviour should be
expected.

In the case (\ref{muthresh}), near the other threshold $4 \mu^2$, the spectrum
for the bound state region $\mu < m$ has the form
\be
E = - \frac{m_{rd} e^8}{2 (l + 1)^4} \left( \frac{2}{\pi} \frac{m^4}{\mu^4}
\sqrt{\frac{\mu^2}{m^2 - \mu^2}} \,  \arctan \sqrt{\frac{\mu^2}{m^2 - \mu^2}}
\, \right)^2 \:, \label{quartic}
\ee
where now $m_{rd} = \mu/2$, provided that $e^4 \ll |m - \mu|/m$. Interestingly
enough, this is qualitatively different from the Coulomb spectrum, and the
complicated expression in parentheses in (\ref{quartic}) provides a
$(m^2/\mu^2)$-dependent modification of the effective coupling constant. In
the case $m^2 \gg \mu^2$, the spectrum becomes
\be
E = - \frac{2 m_{rd}}{\pi^2 (l + 1)^4} \left( \frac{me^2}{\mu} \right)^4 \:,
\ee
indicating that one should rather consider
\be
\frac{me^2}{\mu} = \frac{g^2}{16 \pi m \mu}
\ee
as the effective coupling constant in this case. Finally, there are no
physical states in the ``resonance region'' $\mu > m$, since the imaginary
part of the Regge trajectory is too large to interpret the corresponding poles
of the scattering amplitude as propagating particles.

It will be interesting to further investigate the case of
only slightly different masses $|m - \mu| \ll m$, where the condition
$e^4 \ll |m - \mu|/m$ may or may not be fulfilled. In the latter instance,
the spectrum can be obtained by solving a quartic equation, hence an analytic,
although certainly involved, expression can still be given.
The case $|m - \mu| \ll m$ allows for a
non-relativistic limit for the $\phi$ {\it and}\/ the $\chi$ particles (see
the footnote on page \pageref{nonrelmix}), and so should make a comparison
with the corresponding Bethe-Salpeter equation in its non-relativistic limit
possible. This, in turn, could shed some light on the understanding of the
results found above, as well as provide a description of the ``crossover'' to
the case of equal masses $\mu = m$. We hope to come back to this issue in
the near future.

We have also solved eqs.\ (\ref{spectrum}) for $m_1=m_2=m$, 
(\ref{chzerogeneral}) and the corresponding equation for the charge one sector
numerically, for the cases $m=2\mu$ and $m=\mu/2$ as examples.
The results for the different ($l = 0$) bound states are presented 
in fig.\ 3. 
\begin{figure}
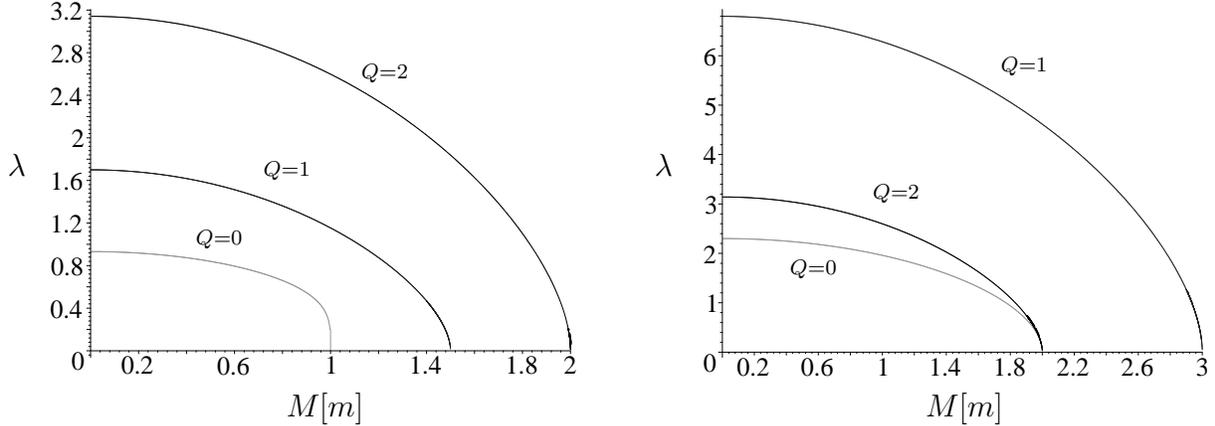
  %%% figure 3
\begin{center}
\begin{picture}(165,61)
%\put(0,0){\framebox(165,61){}}
%%% delta=1
%\put(0,0){\framebox(80,61){}}
\put(4,-17){\psfig{figure=FIGURE3a.ps,height=80mm,angle=-90}}
%\put(29,3){$M [m]$}
\put(38,3){$M [m]$}
\put(1,35){$\lambda$}
\put(26,26){$\scriptstyle Q=0$}
\put(35,35){$\scriptstyle Q=1$}
\put(48,48){$\scriptstyle Q=2$}
%%%  delta = 0.01
%\put(85,0){\framebox(80,61){}}
\put(88,-17){\psfig{figure=FIGURE3b.ps,height=80mm,angle=-90}}
%\put(114,3){$M [m]$}
\put(123,3){$M [m]$}
\put(87,35){$\lambda$}
\put(105,22){$\scriptstyle Q=0$}
\put(116,32){$\scriptstyle Q=2$}
\put(133,49){$\scriptstyle Q=1$}
\end{picture}
\end{center}
\vspace{-6mm}
\caption{Graph of the coupling $\lambda = g^2/(16 \pi)$ versus masses of the 
different ground states in the sectors with charges $Q=0$, 1 and
2 for $m=2\mu$ (left) and $m=\mu/2$ (right), respectively.}
\end{figure}
We see that the charge zero bound states for a given 
coupling have the lowest mass and therefore are the most tightly bound. 
Alternatively, for given constituent masses the charge zero bound state becomes
tachyonic at a smaller coupling than the non-zero charge states. This implies 
that the corresponding vacuum state will be unstable most likely giving rise 
to a vacuum condensate of charge zero bound states. Note that in the case  
$m=\mu/2$ the charge two bound states are less massive than their charge one 
counterparts. Also remark the quartic behaviour of the charge zero bound state 
mass near threshold for $m=2\mu$.

\subsection{Conclusions}

We now have to ask how realistic these results are and what we gain by using
this method? Many of the advantages of the method stem from the fact that
it is intrinsically a quantum field theory methodology and that the
approximations used preserve all relevant symmetries, especially crossing
symmetry,
and in the case of gauge fields --- gauge invariance. Preserving the crossing
symmetry
in particular is crucial in order to obtain certain qualitative features that
otherwise would be lost. 

In the case of $\phi^{\dag}\phi$-states,
where $\mu<m$, an imaginary part to the spectrum was found. Hence, our
methodology gives an integrated method for finding both the real and imaginary
parts
of the spectrum. It is well worth noting that even the fully field theoretic 
Bethe-Salpeter equation does not
reveal this feature in the tree-level ladder approximation, unless one 
uses a matricial equation.
Of course, if one is prepared to consider a one-loop Bethe-Salpeter kernel then
imaginary
parts can be derived without the benefit of a matrix formulation. However, as
is well
known, including loop corrections in such a kernel is far from easy.

Although the decay of unstable states could be described directly using our 
techniques,
the magnitude of the decay rate is very different from that found by a Fermi
Golden Rule treatment. In the latter approach, as mentioned, the bound state 
wave functions play a crucial r\^ole, and there appears to be an effect of
the dimensional reduction in the Regge limit on these wave functions.

A further important advantage of the present method, though one that we have
neither
considered nor exploited here, is its ready extension to higher loop orders.
As mentioned, it is notoriously difficult to solve the Bethe-Salpeter equation
including loop corrections to the kernel. In the present case 
higher loop corrections are included
in a very natural fashion as higher order corrections to the renormalization
constants and their corresponding anomalous dimensions.

We compared our results with those of the Bethe-Salpeter equation for the
Wick-Cut\-kos\-ky
model in the charge two sector. In the relativistic, strong-coupling regime we
see
there are significant differences between our results and those of the
Bethe-Salpeter
equation. It should be emphasized though that, in the absence of a
``benchmark'' result
from an exact solution or an experiment, this in no way invalidates our
approach.
Calculating to two loops will give us a better idea of the size of higher order
corrections. In this sense, even if these corrections are large it may well be 
possible to use a technique
such as Pad\'e resummation in order to obtain more non-perturbative information
such as
is done in calculating fixed points from beta functions. 

The chief disadvantage of calculating bound state properties in terms of Regge
trajectories would seem to be that, at least in the one-loop
approximation, the dependence on the mass of the
``rung'' particles is absent. It is for this reason that the corresponding
spectra
look Coulomb-like rather than Yukawa-like. Of course, this would not be a
problem for a theory where the
relevant exchange particles are massless. Neither does it seem to be a 
problem, as indicated in fig.\ 1, for sufficiently strong coupling.
It remains to be seen whether higher order 
calculations ameliorate this problem.

In the charge one and charge zero sectors we found no corresponding 
Bethe-Salpeter results with which to compare. However, to gain a better
insight into
the physical situation and the significance of our results, we found it 
convenient to consider the non-relativistic limits of the corresponding 
Bethe-Salpeter equations, leading to different types of Schr\"odinger
equations. The upshot of these investigations is that the Regge trajectories
reproduce the qualitative and quantitative features of the situation,
except for their independence of the masses of the exchanged particles.

In the absence of other results, it was interesting to
compare the bound state masses in the different charge sectors with one
another. We saw that, in the cases considered, charge zero bound 
states for a given coupling have the lowest mass and therefore are the most 
tightly bound. A consequence of this is that for given constituent masses the 
uncharged sector becomes
tachyonic at a smaller coupling than the charged sector. This implies 
that the corresponding vacuum state will be unstable most likely giving rise 
to a vacuum condensate of charge zero bound states. 

We may also compare with the Wilsonian RG approach of ref.\ \cite{ellwanger}.
There,
for the Wick-Cutkowsky model, it was not possible to access the
non-relativistic
regime, something which a priori is clearly not a problem in our approach.
Additionally, it would be more difficult to include higher order effects,
such
as the six-point function, in the Wilsonian approach than in the present one
where
two loop effects, though tedious, should be relatively straightforward.

In summary, we have presented a new method for computing bound state properties
based on RG calculations of Regge trajectories. We found a rich bound state
structure in the simple $\phi^{\dag}\phi\chi$ theory considered here. Having
shown the feasibility of such calculations, the task now is to consider 
more realistic theories and evaluate the importance of higher order terms. 
For the latter, it would
be of great interest to compare the results of our approach and those of the
Bethe-Salpeter equation in the context of an exact model where one may see the
deviations from a known result present in the two different methodologies.

\end{fmffile} 
\end{document}